\documentclass[aps,prd,twocolumn,showpacs,floats,floatfix,letterpaper,nofootinbib,superscriptaddress,preprintnumbers]{revtex4}

\usepackage{amssymb,amsmath,latexsym,mathrsfs}
\usepackage{graphicx}
\usepackage{epsfig}
\usepackage{multirow}
\usepackage{array}
\usepackage{xspace}
\usepackage{color}

\newcommand{\bea}{\begin{eqnarray}}
\newcommand{\eea}{\end{eqnarray}}

\newcommand{\neff}{N_{\textrm{eff}}}

\newcommand{\mnu}{{\Sigma}m_{\nu}}

\begin{document}


\title{Cosmological Axion and neutrino mass constraints from Planck 2015 temperature and polarization data}

\author{Eleonora Di Valentino}
\affiliation{Institut d'Astrophysique de Paris (UMR7095: CNRS \& UPMC- Sorbonne Universities), F-75014, Paris, France}
\author{Elena Giusarma}
\affiliation{Physics Department and INFN, Universit\`a di Roma ``La Sapienza'', Ple Aldo Moro 2, 00185, Rome, Italy}
\author{Massimiliano Lattanzi}
\affiliation{Dipartimento di Fisica e Science della Terra, Universit\`a di Ferrara and INFN,\\
sezione di Ferrara, Polo Scientifico e Tecnologico - Edficio C Via Saragat, 1, I-44122 Ferrara Italy}
\author{Olga Mena}
\affiliation{IFIC, Universidad de Valencia-CSIC, 46071, Valencia, Spain}
\author{Alessandro Melchiorri}
\affiliation{Physics Department and INFN, Universit\`a di Roma ``La Sapienza'', Ple Aldo Moro 2, 00185, Rome, Italy}
\author{Joseph Silk}
\affiliation{Institut d'Astrophysique de Paris (UMR7095: CNRS \& UPMC- Sorbonne Universities), F-75014, Paris, France}
\affiliation{AIM-Paris-Saclay, CEA/DSM/IRFU, CNRS, Univ. Paris VII, F-91191 Gif-sur-Yvette, France}
\affiliation{Department of Physics and Astronomy, The Johns Hopkins University Homewood Campus, Baltimore, MD 21218, USA}
\affiliation{BIPAC, Department of Physics, University of Oxford, Keble Road, Oxford
OX1 3RH, UK}
\begin{abstract}
Axions  currently  provide the most compelling solution to the strong CP problem. These particles may be copiously produced in the early universe, including via thermal processes. Therefore, relic axions constitute a hot dark matter component and their masses are strongly degenerate with those of the three active neutrinos, as they leave identical signatures in the different cosmological observables. In addition, thermal axions, while still relativistic states, also contribute to the relativistic degrees of freedom, parameterised via $\neff$. We present the cosmological bounds on the relic axion and neutrino masses, exploiting the full Planck mission data, which include polarization measurements. In the mixed hot dark matter scenario explored here, we find the tightest and more robust constraint to date on the sum of the three active neutrino masses, $\sum m_\nu <0.136$~eV at $95\%$~CL, obtained in the well-known linear perturbation regime. The Planck Sunyaev-Zeldovich cluster number count data further tightens this bound, providing a $95\%$~CL upper limit of $\sum m_\nu <0.126$~eV in this very same mixed hot dark matter model, a value which is very close to the expectations in the inverted hierarchical neutrino mass scenario. Using this same combination of data sets we find the most stringent bound to date on the thermal axion mass, $m_a<0.529$~eV at $95\%$~CL.
\end{abstract}
\preprint{IFIC/15-60}

\maketitle

{\it \textbf{Introduction ---}} The axion field arises as a solution to solve the strong CP problem in Quantum Chromodynamics~\cite{PecceiQuinn,Weinberg:1977ma,Wilczek:1977pj}. The axion is  the Pseudo-Nambu-Goldstone associated with a new global $U(1)_{PQ}$ (Peccei-Quinn) symmetry that is spontaneously broken at an energy scale $f_a$. In the early universe, axions can be produced via thermal or non-thermal processes.  While in the former case, the axion contributes an extra hot thermal relic (together with three active sterile neutrinos), in the latter case, the axion could be the cold dark matter component. In the following, we shall focus on the thermal axion scenario. In order to compute the present thermal axion relic density, the most relevant process is the axion-pion interaction, $ \pi + \pi \rightarrow \pi+a$.
The characteristic parameter for the thermal axion is $f_a$, the axion coupling constant, that can be related to the axion mass by
\begin{equation}
m_a = \frac{f_\pi m_\pi}{  f_a  } \frac{\sqrt{R}}{1 + R}=
0.6\ {\rm eV}\ \frac{10^7\, {\rm GeV}}{f_a}~,
\end{equation}
where the up-to-down quark mass
ratio is  taken as $R=0.553 \pm 0.043 $, and $f_\pi = 93$ MeV is the pion decay constant. 

Thermal axions, while still relativistic, will increase the amount of radiation in the universe, contributing to the effective number of relativistic degrees of freedom $\neff$. In the standard cosmological $\Lambda$CDM model with three active neutrino species, we expect $\neff=3.046$ \cite{Mangano:2005cc}, where the $0.046$ takes into account corrections for the non-instantaneous neutrino decoupling from the primordial plasma. An extra $\Delta \neff = \neff - 3.046$ modifies the damping tail of the Cosmic Microwave Background (CMB) temperature angular power spectrum, changing two important scales at recombination, the sound horizon and the Silk damping scales, as well as  the primordial abundances of the light elements predicted by Big Bang Nucleosynthesis.
When thermal axions become non-relativistic particles, they will affect the different cosmological observables in an analogous way to that  of massive neutrinos, i.e. by increasing the amount of the (hot) dark matter density in our universe. Axions will suppress  structure formation at scales smaller than the free-streaming scale, favouring clustering only at large scales. Thermal axions will also leave an imprint  on  the CMB temperature anisotropies, via the early integrated Sachs-Wolfe effect. Therefore,  a large degeneracy between the axion mass and the total neutrino mass is expected~\cite{Melchiorri:2007cd}. Several papers in the literature have provided cosmological constraints on the thermal axion mass in different cosmological scenarios, see e.g. Refs.~\cite{Melchiorri:2007cd,Hannestad:2007dd,Hannestad:2008js,Hannestad:2010yi,Archidiacono:2013cha,Giusarma:2014zza,DiValentino:2015zta}. 

In light of the recent Planck 2015 temperature and polarization data \cite{Adam:2015rua}, it is timely to compute the changes in the existing bounds on the thermal axion mass, including the case in which massive neutrinos are present. Our results are obtained using the Monte Carlo Markov Chains (MCMC) package \texttt{CosmoMC} \cite{Lewis:2002ah}, with \texttt{CAMB} (Code for Anisotropies in the Microwave Background) \cite{Lewis:1999bs} as a solver of the Boltzmann equations. In the mixed hot dark matter scenario, in which both axion and neutrino masses are  allowed to freely vary, we find the tightest and most robust constraint to date on the sum of the three active neutrino masses, $\sum m_\nu <0.136$~eV at $95\%$~CL, as this only relies  on the (very well-known) linear perturbation regime. 

{\it \textbf{Thermal axion cosmological model}} The scenario we analyze here is the $\Lambda$CDM model, with both axions and neutrinos as extra hot thermal relics. We describe this scenario by the following set of parameters:
\begin{equation}\label{parameter}
\{\omega_b,\omega_c, \Theta_s, \tau, m_a, \sum m_\nu, n_s, \log[10^{10}A_{s}]\}~,
\end{equation}
where $\omega_b\equiv\Omega_bh^{2}$ is the baryon matter energy density, $\omega_c\equiv\Omega_ch^{2}$  the cold dark matter energy density,
$\Theta_{s}$ is the ratio between the sound horizon and the angular diameter distance at decoupling, $\tau$ is the reionization optical depth, $m_a$ is the axion mass in eV and $\sum m_\nu$ the sum of three active neutrino masses in eV. We consider also the inflationary parameters, the scalar spectral index $n_s$ and the amplitude of the primordial spectrum $A_{s}$. We use flat priors for all the parameters, as listed in Tab.~\ref{tab:priors}. Notice that the standard extra radiation density will change, as the presence of a thermal axion will increase the value of the effective number of relativistic degrees of freedom in the following way:
\begin{equation}
\Delta \neff =\frac{ 4}{7}\left(\frac{3}{2}\frac{n_a}{n_\nu}\right)^{4/3}~,
\end{equation}
where $n_a$ is the axion number density and $n_\nu$ is the present neutrino plus antineutrino number density per flavor. The current axion number density is a function of the axion decoupling temperature $T_D$, that is a function of the axion mass $m_a$. For the details related to the calculation of the axion decoupling temperature, we refer the reader to Ref.~\cite{Giusarma:2014zza}. 

\begin{table}
\begin{center}
\begin{tabular}{c|c}
Parameter                    & Prior\\
\hline
$\Omega_{\rm b} h^2$         & $[0.005,0.1]$\\
$\Omega_{\rm cdm} h^2$       & $[0.001,0.99]$\\
$\Theta_{\rm s}$             & $[0.5,10]$\\
$\tau$                       & $[0.01,0.8]$\\
$m_a$ (eV)                        & $[0.1,3]$\\
$\sum m_\nu$ (eV)               & $[0.06,3]$\\
$n_s$                        & $[0.9, 1.1]$\\
$\log[10^{10}A_{s}]$         & $[2.7,4]$\\
\end{tabular}
\end{center}
\caption{
Priors for the parameters used in the MCMC analyses.
}
\label{tab:priors}
\end{table}

{\it \textbf{Datasets}} 
Our baseline data set consists of the recent Planck 2015 satellite CMB temperature and polarization measurements~\cite{Adam:2015rua,Ade:2015xua,Aghanim:2015wva}.  We consider a combination of the likelihood at $30\le \ell\le 2500$ using  TT, TE and EE power spectra and the Planck  low-$\ell$ multipole likelihood in the range $2\le\ell\le29$. We refer to this combination as \emph{Planck  TT,TE,EE+lowP}, following the nomenclature
of Ref. \cite{Ade:2015xua}. We also include the new Planck 2015 lensing likelihood~\cite{Ade:2015zua}, constructed from measurements of the power spectrum of the lensing potential, referring to it as $lensing$. Concerning Planck catalogs, we make use of the Sunyaev-Zeldovich second cluster catalog~\cite{Ade:2015gva,Ade:2015fva} (denoted as \emph{SZ} in what follows), which consists of  439 clusters with their corresponding redshifts and with a signal-to-noise ratio $q>6$. We also consider additional datasets to the Planck satellite measurements, as a gaussian prior on the Hubble constant $H_0=73.8\pm2.4$ km/s/Mpc, according with the measurements of the Hubble Space Telescope,~\cite{Riess:2011yx}. We refer to this data set as \textit{HST}.  We also include measurements of the large scale structure of the universe in their geometrical form, i.e. in the form of Baryon Acoustic Oscillations (BAO). In particular, we use the 6dFGS, SDSS-MGS and BOSS DR11 measurements of $D_V/r_d^2$~\cite{Beutler:2011hx,Ross:2014qpa,Anderson:2013zyy}, referring to the combination of all of them as \textit{BAO}. We shall also consider large  scale structure measurements in their full matter power spectrum form, as provided by WiggleZ  survey~\cite{Parkinson:2012vd}, and denoted as \textit{MPK}. Tomographic weak lensing surveys provide a powerful tool to constrain the mass distribution in the universe, and therefore we shall also exploit in our analyses the constraint on the relationship between $\sigma_8$ and $\Omega_m$ of $\sigma_8(\Omega_m/0.27)^{0.46}=0.774\pm0.040$ provided by  the  Canada-France-Hawaii Telescope~\cite{Heymans:2013fya}, CFHTLenS. This last measurement  is referred to as \textit{WL}.

\begin{table*}
\begin{center}\footnotesize
\scalebox{0.83}{\begin{tabular}{c|cccccccc}
\hline \hline
  & TT,TE,EE+lowP&TT,TE,EE+lowP & TT,TE,EE+lowP & TT,TE,EE+lowP & TT,TE,EE+lowP & TT,TE,EE+lowP & TT,TE,EE+lowP & TT,TE,EE+lowP      \\                             
                                          &                                                                        &+lensing                                      &+WL
&+MPK & +BAO  & +HST  & +BAO +HST & +BAO +HST +SZ \\ 
\hline
\hspace{1mm}\\

$\Omega_{\textrm{c}}h^2$    &$0.1235_{-0.0036}^{+0.0034}$      &$0.1235_{-0.0034}^{+0.0034}$   &$0.1225_{-0.0032}^{+0.0032}$
&$0.1237_{-0.0031}^{+0.0034}$  &$0.1223^{+0.0023}_{-0.0023} $      &$0.1223^{+0.0032}_{-0.0032}$ &$0.1220^{+0.0024}_{-0.0023}$ &$0.1216^{+0.0023}_{-0.0023}$\\
\hspace{1mm}\\

$m_a$ [eV]                           &$<2.09$                                           &$<1.67$                                       &$<1.87$
&$<0.835$ &$<0.763$  &$<1.21$ &$<0.709$&$<0.529$\\
\hspace{1mm}\\

$\sum m_\nu$ [eV]               &$<0.441$                                         &$<0.538$                                     &$<0.360$
&$<0.291$  &$<0.159$  &$<0.182$ &$<0.136$ &$<0.126$\\
\hspace{1mm}\\

$\sigma_8$                          &$0.779_{-0.094}^{+0.083}$            &$0.767_{-0.072}^{+0.065}$         &$0.789_{-0.096}^{+0.074}$
&$0.814_{-0.056}^{+0.049 }$ &$0.827^{+0.039}_{-0.042}$ &$0.820^{+0.051}_{-0.062}$ &$0.829^{+0.036}_{-0.039}$ &$0.835^{+0.033}_{-0.035}$\\
\hspace{1mm}\\                                                                                                                                                                                               
                                                                                                                                                                                                             
$\Omega_{\textrm{m}}$      &$0.342_{-0.048}^{+0.054}$             &$0.344_{-0.048}^{+0.055}$          &$0.328_{-0.041}^{+0.048}$
&$0.326_{ -0.029}^{+0.033}$   &$ 0.312^{+0.016}_{-0.014}$ &$0.315^{+0.031}_{-0.027}$ &$0.309^{+0.015}_{-0.014}$ &$0.306^{+0.014}_{-0.013}$ \\
\hspace{1mm}\\                                                                        
                        
$\log[10^{10} A_s]$            &$3.131_{-0.070}^{+0.067}$               &$3.109_{-0.062}^{+0.064}$          &$3.117_{-0.068}^{+0.071}$
&$3.121_{-0.071}^{+0.066}$    &$3.126^{+0.066}_{-0.070} $ &$3.129^{+0.066}_{-0.068}$ &$3.128^{+0.065}_{-0.069}$ &$3.132^{+0.063}_{-0.064}$\\
\hspace{1mm}\\           
                         
$n_s$                                 &$0.972_{-0.012}^{+0.011}$                &$0.972_{-0.011}^{+0.010}$           &$0.974_{-0.012}^{+0.011}$
&$0.97278_{-0.009}^{+0.009}$    &$0.9754^{+0.0093}_{-0.0089}$ &$0.976^{+0.010}_{-0.010}$ &$0.9763^{+0.0095}_{-0.0091}$ &$0.9768^{+0.0089}_{-0.0089}$\\
\hspace{1mm}\\  
\hline
\hline
\end{tabular}}
\caption{$95\%$~CL constraints on the parameters of the mixed hot dark matter scenario explored here (the $\Lambda$CDM+$m_a$+$\mnu$ model) for the different combinations of cosmological data sets.}
\label{tab:lcdm+ma}
\end{center}
\end{table*}

\begin{figure}[!t]
\includegraphics[width=9cm]{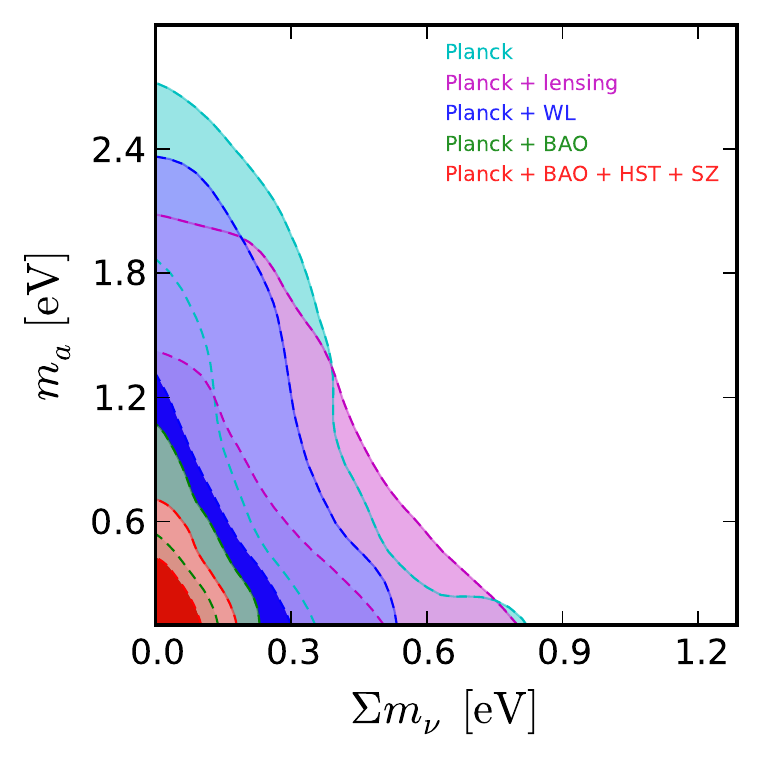}
\caption{$68\%$ and $95\%$~CL allowed regions in the ($\sum m_\nu$, $m_a$) plane, both in eV, for some of the cosmological data combinations explored in this analysis.}
\label{fig:mamnu}
\end{figure}

{\it \textbf{Results}} 
Table \ref{tab:lcdm+ma} summarises the results from our MCMC analyses in the mixed hot dark matter scenario revisited here. Notice that Planck temperature and polarization measurements (\emph{TT,TE,EE+lowP}) set $95\%$~CL upper bounds of $\sum m_\nu<0.441$~eV and $m_a<2.09$~eV respectively. The bounds on the thermal axion mass are similar to those obtained in the case in which only axion masses are considered, albeit for that case  the value of the $\sigma_8$ parameter is always higher than the one shown here, as only one hot relic suppresses the small-scale clustering. Nevertheless the deviation of $\sigma_8$ is not significant (about a half-sigma away from the value illustrated in Tab.~\ref{tab:lcdm+ma}). Furthermore, neutrino oscillation experiments have provided compelling evidence for the existence of neutrino masses and therefore neutrinos must be added as massive particles. The addition of CMB lensing measurements from the Planck satellite weakens the neutrino mass bounds, as discussed in \cite{Ade:2015xua}: the lensing reconstruction data  prefers lensing amplitudes lower than the standard prediction, and this favours higher neutrino masses, as the presence of these will smooth the lensing power spectrum.  Summarizing, when Planck CMB lensing constraints are considered, the neutrino mass bounds are pulled away from zero, and we obtain $\sum m_\nu<0.538$~eV and $m_a<1.67$~eV  at $95\%$~CL. 
The addition of weak lensing constraints on the relationship between the matter clustering amplitude $\sigma_8$ and the matter mass-energy density $\Omega_m$ to \emph{Planck TT,TE,EE+lowP} measurements only mildly tightens both the thermal neutrino and axion masses. The largest impact on both $\sum m_\nu$ and $m_a$ bounds comes from the large-scale structure information as well as from the prior on $H_0$ from the \emph{HST} experiment. Notice that the bounds are significantly tighter when one of the former constraints are considered in the analyses. Concerning the $H_0$ prior, the $95\%$~CL upper bounds on the thermal relic masses become $\sum m_\nu<0.182$~eV and $m_a<1.21$~eV. The reason for this large improvement is due to the large degeneracy between $\sum m_\nu$ and $H_0$\cite{Giusarma:2012ph}. When $\sum m_\nu$ increases there is a shift in the distance to last scattering. This shift can be easily compensated by lowering $H_0$, resulting in a strong degeneracy between these two parameters, which can be broken via an independent measurement of  $H_0$. However, the tightest axion and neutrino mass constraints arise when large-scale structure data is exploited in its geometrical form, via the BAO signature. Indeed, it was shown in Ref.~\cite{Hamann:2010pw} that, when constraining $\sum m_\nu$ in minimal schemes as the one explored here (i.e. a $\Lambda$CDM model), the information contained in the broadband shape of the halo power spectrum was superseded by geometric information derived from the BAO signature. We find here a similar effect, although the BAO measurements that we exploit correspond to several redshifts and surveys, while the full-shape data come from only one survey, the WiggleZ survey. Using the full matter power spectrum measurements from the former experiment, we obtain $95\%$~CL upper bounds of $\sum m_\nu<0.291$~eV and $m_a<0.835$~eV. The $95\%$~CL upper bound of $\sum m_\nu< 0.159$~eV for the \emph{Planck TT,TE,EE+lowP} and \emph{BAO} combination is very close to the one quoted by the Planck collaboration for the same data sets, $\sum m_\nu< 0.17$~eV~\cite{Ade:2015xua}. However, our constraint is tighter, as we are also considering here axions as additional thermal relics, and there exists a strong degeneracy among these $\sum m_\nu$ and $m_a$. Figure \ref{fig:mamnu} illustrates such a degeneracy. We depict, in the ($\sum m_\nu$, $m_a$) plane, the $68\%$ and $95\%$~CL contours arising from the analyses of \emph{Planck TT,TE,EE+lowP} data plus additional measurements, as the Planck lensing signal and other data sets (\emph{WL}, \emph{BAO}, \emph{HST} and \emph{SZ} cluster number counts). Notice that the constraints are greatly improved for the former two cases, leading to very tight constraints on the masses of these two thermal relics. 

 The addition of the \emph{BAO}  and the \emph{HST} data sets leads to the strongest constraint on the neutrino mass to date in the linear perturbation regime, $\sum m_\nu< 0.136$~eV at $95\%$~CL. The corresponding bound on the axion mass is $m_a<0.709$~eV. The authors of \cite{Palanque-Delabrouille:2015pga} have recently reported, using the one-dimensional Lyman-$\alpha$ forest power spectrum of the BOSS experiment, a $95\%$~CL upper bound of $\sum m_\nu< 0.12$~eV in the case in which only  massive neutrinos are present. Notice however that this constraint strongly relies on hydrodynamical simulations, while our bounds are derived in the very well-known linear perturbation regime. Furthermore, the addition of the Planck \emph{SZ} cluster number counts data provide a competitive $95\%$~CL upper limit  of $\sum m_\nu <0.126$~eV in the mixed axion-neutrino hot dark matter scenario (the corresponding bound on the thermal axion mass is $m_a <0.529$~eV). This limit is very close to the expectations for $\sum m_\nu$  in the inverted hierarchical neutrino mass scenario, highlighting the fact that improved cluster mass calibrations could help enormously in disentangling the neutrino mass spectrum.

{\it \textbf{Conclusions}} The polarization measurements from the Planck 2015 data release offer a unique opportunity for testing the dark matter paradigm. These recent results point to a standard $\Lambda$CDM as the preferred model for the universe we observe today. Nevertheless, a small hot dark matter component can still be present. We have explored the most general scenario, i.e. a mixed hot dark matter model with two thermal relics, neutrinos and axions, which would account for the small contribution from the hot dark matter sector to the total mass-energy density of the universe. Using Planck temperature and polarization data, and making use of the Planck Sunyaev-Zeldovich cluster catalog as well as independent, low redshift probes, including measurements of the Baryon Acoustic peak in galaxy clustering and  of the Hubble constant, we derive the tightest bounds to date on the thermal relic masses. The $95\%$ upper limits extracted from the numerical analyses carried out in this study are $m_a <0.529$~eV and $\sum m_\nu <0.126$~eV for the axion and total neutrino mass, respectively. These results strongly motivate the need for improved cluster mass calibrations. They also clearly illustrate the power of combining low and high redshift probes when cornering the dark matter thermal properties. 

{\it \textbf{Acknowledgments}}  OM is supported by PROMETEO II/2014/050, by the Spanish Grant FPA2011--29678 of the MINECO and by PITN-GA-2011-289442-INVISIBLES. This work has been done within the Labex ILP (reference ANR-10-LABX-63) part of the Idex SUPER, and received financial state aid managed by the Agence Nationale de la Recherche, as part of the programme Investissements d'avenir under the reference ANR-11-IDEX-0004-02. EDV acknowledges the support of the European Research Council via the Grant  number 267117 (DARK, P.I. Joseph Silk).



\end{document}